\documentclass[%
 reprint,
 amsmath,amssymb,
 aps,
 prl,
]{revtex4-2}

\usepackage{graphicx}% Include figure files
\usepackage{dcolumn}% Align table columns on decimal point
\usepackage{bm}% bold math
\usepackage{xcolor}
\usepackage{hyperref}% add hypertext capabilities
\usepackage[normalem]{ulem}
\usepackage{stackrel}
\usepackage{caption}
\usepackage{subcaption}
\usepackage{braket}
\usepackage{float}

\begin{document}

\preprint{APS/123-QED}

\title{Neural Networks as Universal Probes of Many-Body Localization in Quantum Graphs}

\author{Cameron Beetar}
 \email{btrcam001@myuct.ac.za}
 
\author{Jeff Murugan}%
 \email{Jeff.Murugan@uct.ac.za}
\affiliation{%
 The Laboratory for Quantum Gravity \& Strings,\\
 Department of Mathematics \& Applied Mathematics,\\
 University of Cape Town, Rondebosch, Cape Town, South Africa
 }%
\author{Dario Rosa}
\email{dario\_rosa@ibs.re.kr}
\affiliation{%
 Center for Theoretical Physics of Complex Systems,\\
Institute for Basic Science (IBS),\\
Daejeon - 34126, Korea
}%

\date{\today}% It is always \today, today,

\begin{abstract}
We show that a neural network, trained on the entanglement spectra of a nearest neighbor Heisenberg chain in a random transverse magnetic field, can be used to efficiently study the ergodic/many-body localized properties of a number of other quantum systems, without further re-training. We benchmark our computational architecture against a $J_{1}\!\!-\!\!J_{2}$ - model, which extends the Heisenberg chain to include next-to-nearest neighbor interactions, with excellent agreement with known results. When applied to Hamiltonians that differ from the training model by the topology of the underlying graph, the neural network is able to predict the critical disorders and, more generally, shapes of the mobility edge that agree with our heuristic expectation. We take this as proof of principle that machine learning algorithms furnish a powerful new set of computational tools to explore the many-body physics frontier.     
\end{abstract}

\maketitle

%\tableofcontents

%\section{\label{sec:intro}Introduction}
\emph{I. Introduction.---\label{sec:intro}}
Science covets universal tools. Calculus, statistical data analysis and perturbation theory all fall into this category - powerful conceptional and computational frameworks applicable across a wide spectrum of disciplines. The advent of computers, and scientific computing in particular, was another paradigm shift which reshaped the research landscape since the 1980's. Today it is evident that we stand on the cusp of another such revolution, as {\it machine learning} algorithms begin to make an impact on nearly every facet of modern scientific research, from medical diagnostics to mathematical theorem proving. The seemingly universal applicability, and remarkable adaptability, of such algorithms makes it an ideal computational tool to address problems at the frontier of numerical (and analytic) accessibility in physics.

Many-body quantum physics is replete with such problems. Among these, the localization properties of disordered quantum matter stand out, as much for their complexity as for their broad relevance. It is for these precise reasons, armed with new tools from machine learning, that we return to this problem in the current article. This is, of course, not the first time that neural networks have been turned loose on the problem of many-body localization (MBL). In fact, our strategy for attacking this problem is based largely on \cite{PhysRevB.95.245134} in which the thermalization to localization transition - the so-called {\it mobility edge} - was computed for a disordered, nearest-neighbour Heisenberg spin chain, as a function of a random magnetic field $\bar{h}$. The computation of the mobility edge even in a toy model like the Heisenberg chain is computationally demanding, requiring sophisticated numerical tools, large system sizes (up to 24 sites) and many disorder realizations \cite{PhysRevB.91.081103, PhysRevLett.125.156601}. In comparison, even a single hidden-layer neural network trained on the entanglement spectrum of the Heisenberg chain is able to reproduce all the qualitative features of the MBL phase diagram with significantly smaller system sizes (a 16-site spin chain) and substantially fewer disorder realizations \cite{PhysRevB.95.245134}.

In this article, we will push this analysis even further and ask whether a neural network, trained on the entanglement spectra of the same (nearest-neighbour) disordered Heisenberg spin chain as in \cite{PhysRevB.95.245134}, is able to predict the mobility edge for a wide range of Hamiltonians defined on topologically different networks, {\it without additional re-training}. The quantum networks that we consider include (i) next-to-nearest neighbour interactions whose graph exhibits clustering, (ii) star graphs with a very small interaction length and (iii) a superposition of nearest neighbour and star interactions whose graph, with both simple clustering as well as a small diameter, is an example of a scale-free network which is a special case of a small-world network \cite{watts_collective_1998}. This is an (albeit limited) example of what is called {\it transfer-learning} in the machine-learning literature - think of a child, having learnt how to walk adapting those mechanical movements to learn how to run, again, without having to re-train on a different set of movements. 

In what follows, we will set up the physics problem we would like to study - the disordered spin chain defined on quantum networks - in sections {\bf II} and {\bf III}. In section {\bf IV} we describe briefly the neural network architecture that we use to probe the quantum systems in question, with our results described in section {\bf V}. Further details of the neural network, as well as comparisons of our results to previous work can be found in the supplementary material. \\

%\section{\label{sec:Heisenberg}The ETH/MBL transition}
\emph{\label{sec:Heisenberg}II. The ETH/MBL transition.---}
Otherwise thermalizing systems can sometimes exhibit properties of localization due to disorder \cite{doi:10.1146/annurev-conmatphys-031214-014726, ALET2018498, RevModPhys.91.021001}.
The standard example for this phenomenon, by now well-understood and accepted, is the Anderson model --- a non-interacting free fermion model hopping on a lattice and subject to an on-site, random, potential \cite{PhysRev.109.1492}.
Depending on the strength of the disorder, the energy eigenstates can be extended (for small enough disorder) or localized, \textit{i.e.} having non-vanishing support only on few sites. 
%The latter case is the phenomenon of localization, triggered by the disordered potential.
The disorder strength at which the transition from extended to localized behavior happens is usually called \textit{critical disorder} and it is in general highly dependent on the dimensionality on which the model is defined \cite{doi:10.1142/7663}.

Many-body localization addresses the fate of the localized phase in presence of interactions \cite{PhysRevLett.95.206603, BASKO20061126, PhysRevB.76.052203}.
In this case, the problem becomes immediately much more involved. 
For example, from a computational point of view, one has to deal with an exponentially large (in the number of particles, $N$) Hilbert space; the latter being an intrinsic property of any quantum many-body system.
Consequently, any exact numerical study is constrained to deal with systems having just few degrees of freedom \cite{PhysRevB.75.155111, PhysRevB.77.064426, PhysRevB.82.174411, PhysRevB.91.081103, PhysRevLett.125.156601}, thus making the extrapolation of the finite $N$ results to the thermodynamic limit a formidable task, as well as a matter of ongoing debate \cite{PhysRevE.102.062144, ABANIN2021168415, PhysRevB.102.064207, PhysRevB.102.014201, Panda_2020, PhysRevB.102.100202, PhysRevB.102.125134, morningstar2021avalanches, PhysRevB.93.014203, PhysRevB.102.060202}.

Independently of the behavior in the thermodynamic limit, many finite $N$ studies of MBL systems have been already performed in the past.
Among the models which have been studied, the Heisenberg model in a random static magnetic field has by now become the ``gold standard'' example of an MBL system and, for this reason, this is the first model we discuss and it will constitute the model that we will use to train our neural network.
This step will allow us to perform some sanity checks on our results against the amount of results already at disposal in the literature, including the results of \cite{PhysRevB.95.245134} which uses the same neural network architecture.

The $N$-site Heisenberg model describes a set of $N$ spin-1/2 particles, living on  a chain with periodic boundary conditions, in presence of a random static transverse magnetic field. The model is  described by the Hamiltonian \cite{PhysRevB.82.174411}
\begin{equation}
    H = \frac{1}{4}\sum_{i, j} J_{ij} \bm{\sigma}_i\cdot\bm{\sigma}_j+\frac{1}{2}\sum_{i=1}^N h_i\sigma^z_i,
    \label{eq:heis}
\end{equation}
where the $\sigma^z_i$ corresponds to the $z$-component Pauli matrix at the $i^{\text{th}}$ site, the $\bm{\sigma}_i\cdot\bm{\sigma}_j$ is the scalar product between the Pauli matrices at the $i^{\text{th}}$ and $j^\text{th}$ sites and the $h_i$ are the random static magnetic field values, pseudo-randomly sampled from a uniform distribution $h_i\in\left[-W,W\right]$, $W\in\mathbb{R}^+$ being the disorder strength.  In the Hamiltonian above, the interaction coupling matrix $J_{ij}$ can be thought of as the adjacency matrix for a nearest-neighbour graph, on which the model lives, and so it is defined to be $1$ when $i$ and $j$ are adjacent and $0$ otherwise ({\it i.e.} $J_{ij}\to J\delta_{i,i+1}$).
Since we have implicitly set $\hbar=1$, we measure energy in units of $J\hbar$. 
As already mentioned, this model has been extensively studied in recent years.
In particular, while exactly solvable for $W = 0$, it turns out to satisfy ETH at small disorder, while the eigenstates become localized at large values of the disorder.
The value of the disorder at which we measure the transition from extended to localized behavior turns out to be energy-dependent, a feature that defines the presence of a mobility edge \cite{doi:10.1142/7663, PhysRevB.91.081103}. This means that, at intermediate disorder strengths, regions of the spectrum in the extended regime coexist with other regions (at different energies) in the localized regime.
Given that the model, at least at finite $N$, is by now well-understood, it constitutes a perfect playground in which to train our neural network using a supervised approach. Details of the training algorithm as well as a comparison to existing predictions for the phase diagram can be found in the Appendix.\\

% \section{\label{sec:Graphs}Rewiring the Interactions}
\emph{\label{sec:Graphs}III. Rewiring the Interactions.---}
Once the network is trained, and results benchmarked against known results for the nearest-neighbour Heisenberg chain, we turn our attention to the real problem of interest: \textit{systems with different entanglement topologies as coded in the underlying interaction graph}. In particular, we will consider three classes of graphs, each encoding a specific property that we would like to explore (see Figure 1):

The {\bf next-to-nearest neighbour} Hamiltonian is specified by setting $J_{ij} = J_{1}\,\delta_{i,i+1}+J_{2}\,\delta_{i,i+2}$ and introduces two qualitatively new properties into the system\footnote{In the special case $J_{1}=2J_{2}$ and $W=0$, this becomes the celebrated and exactly solvable Majumdar-Ghosh model \cite{1969JMP....10.1388M} for which the Hamiltonian is directly expressed in terms of the quadratic Casimir of the 3-site spin algebra.}; longer-range interactions and {\it clustering}. This latter property is a feature of any graph with a large fraction of triangles present and can be quantified by a clustering coefficient which, at site $i$ is counted as $C_{i} = 2\Delta_{i}/(k_{i}(k_{1}-1))$, if the site had degree $k_{i}$ and has $\Delta_{i}$ triangles attached to it. The corresponding graph-averaged clustering coefficient is then $\overline{C} = \sum_{i} C_{i}/N$. 
In addition, numerical studies for the ETH/MBL transition for this model have been already obtained \cite{PhysRevE.102.062144}, thus making it as a control case to check the performance of the NN on new models. 

The $N$-site {\bf star graph} in which one site, say $\alpha$, is singled out and connected to each of the remaining $N-1$ sites with no other edges, so that $J_{ij} = J_{\alpha} \delta_{\alpha i}$. The star graph is an example of a complete bipartite graph with paths in the graph having either length 1 (for sites connecting to $\alpha$) or 2 (for sites connected via $\alpha$). The resulting average distance in the network $\bar{l} = (1/N)\sum_{i}\bar{l}_{i} = 2 - 4/N + 3/N^{2}$ is exact, and characteristically small.

The {\bf bicycle-wheel graph} defined by the adjacency matrix $J_{ij} = J_{1} \delta_{i\neq\alpha,i+1} + J_{3}\delta_{\alpha i}$ superposes the nearest-neighbour Hamiltonian \eqref{eq:heis} with the star graph and is actually a simple example of a broader class of {\it small-world} graphs \cite{watts_collective_1998} that exhibit both clustering (due to the graph transitivity) as well as a very short average path length (as a result of shortcuts through the hub). Here again, the graph, as a representative of its class, is sufficiently simple that the network characteristics can be computed exactly. Specifically, the average clustering coefficient is $\overline{C} = 2/3 - 2/3N + 2/(N^{2}-N)$, while the average path length reads $\bar{l} = 2 - 6/N + 5/N^{2}$.

%This is done by altering the first sum in %\eqref{eq:heis} to moderate which sites may interact. %The Hamiltonian \eqref{eq:heis} is modified to be
%\begin{align}
%    H_{\text{NNN}} &= \frac{J_1}{4}\sum_{\langle i %j\rangle} \bm{\sigma}_i\cdot\bm{\sigma}_j+ %\frac{J_2}{4}\sum_{\langle\langle i j %\rangle\rangle}\bm{\sigma}_i\cdot\bm{\sigma}_j+\frac{1%}{2}\sum_{i=1}^N h_i\sigma_i^z \label{eq:NNN},\\
%    H_{\text{NNS}} %&=\frac{J_1}{4}\sum_{\stackrel{i,j\neq \alpha}{\langle %i j\rangle}} \bm{\sigma}_i\cdot\bm{\sigma}_j + %\frac{J_3}{4}\sum_{i\neq %\alpha}\bm{\sigma}_\alpha\cdot\bm{\sigma}_i %+\frac{1}{2}\sum_{i=1}^N h_i\sigma_i^z %\label{eq:NNStar},\\
%    H_{\text{Star}} &= \frac{J_3}{4}\sum_{i\neq %\alpha}\bm{\sigma}_\alpha\cdot\bm{\sigma}_i %+\frac{1}{2}\sum_{i=1}^N h_i\sigma_i^z &\label{eq:Star},
%\end{align}

In each case, we will retain the random static transverse magnetic field term $\sim\sum h_{i}\sigma_{i}^{z}$ that disorders the system. In this way, we are able to investigate the competing effects of disorder which tends to promote many-body localization and interactions which favour thermalization while keeping the system size $N$ within a computationally managable range.\\

\begin{figure}
    \centering
    \includegraphics[scale = 0.16]{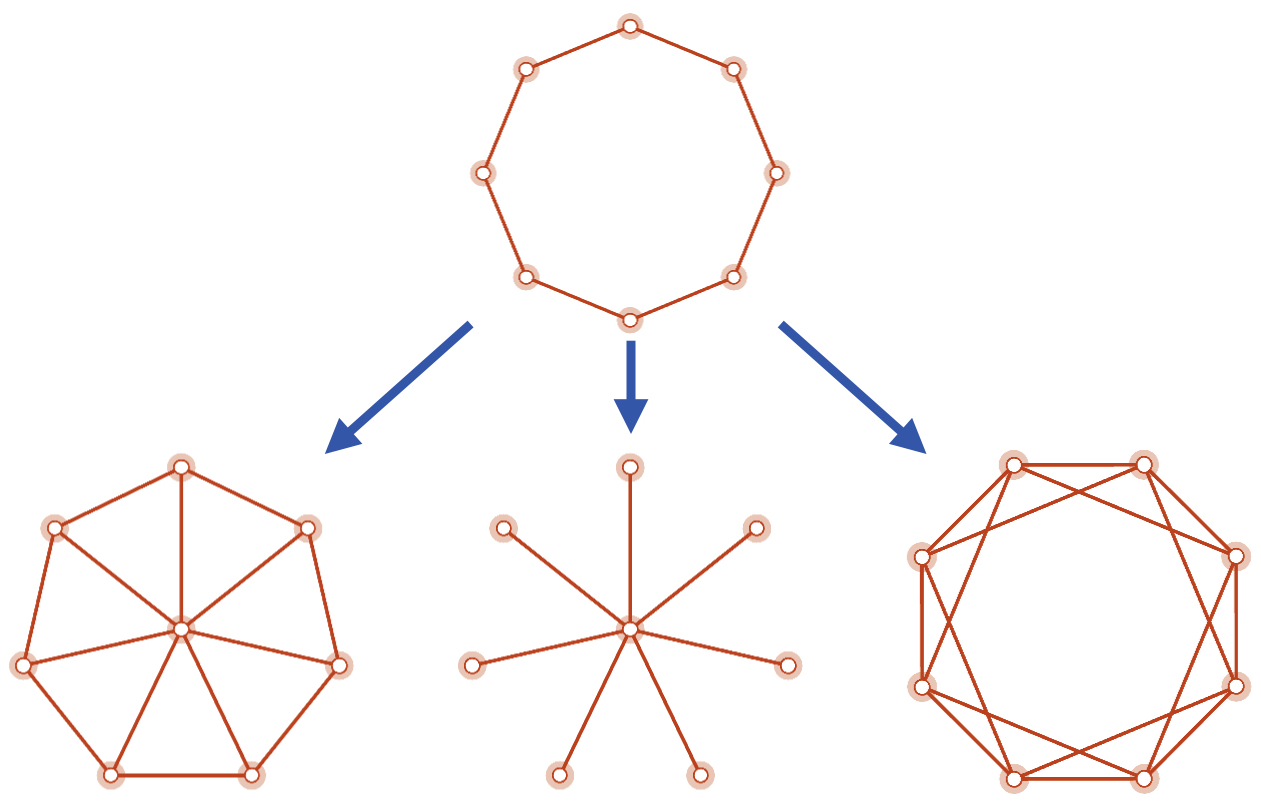}
    \captionsetup{justification=raggedright}
    \caption{Interaction graphs for an 8-site system. Top: Nearest-neighbour interactions on which the neural network is trained. Bottom, from left to right: superposition of star and nearest-neighbour interactions; a star graph and next to nearest-neighbour circulate graph. The central nodes in the lower-left images corresponds to the $\alpha$-site in the defining adjacency matrices.}
    \label{fig:Topologies}
\end{figure}

%\section{\label{sec:NN}Binary Classification Neural Network}
\emph{\label{sec:NN}IV. Binary Classification Neural Network.---}
Distinguishing between ETH and MBL regimes in these many-body systems is essentially a classification problem. As such, the mobility edge that forms the  boundary between these two phases corresponds to a band of classification uncertainty in the algorithm. To see how this works, we will adopt a binary classification neural network scheme similar to that in \cite{PhysRevB.95.245134}. In short, the neural network maps an $n$-dimensional vector, consisting of the entanglement spectrum at a given eigenenergy and a specified disorder, to a 2-dimensional vector that quantifies how confident the network is that the entanglement spectrum corresponds to an ETH or localizing phase. We outline here some of the key features of our neural network, and refer the reader to the Supplementary Material for further elaboration.

\begin{figure}[h]
    \centering
    \includegraphics[scale=0.7]{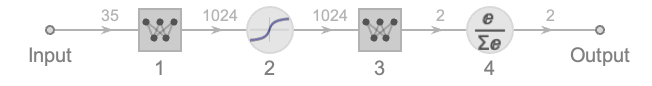}
    \captionsetup{justification=raggedright}
    \caption{A visualisation of the neural network, and its layers, associated with $N=16$ systems. Numbers above the directed edges correspond to the number of neurons between each transformation. Icons labelled 1 and 3 correspond to linear transformations while icons labelled 2 and 4 correspond to nonlinear activation functions.}
    \label{fig:NetChain}
\end{figure}

A defining feature of artificial neural networks (ANN) are their layers. Mathematically speaking, a \textit{layer} consists of a linear map and a nonlinear \textit{activation function}. The layer is called \textit{hidden} if it is not the final layer of the network. For the purposes of this work, we use a single hidden layer ANN (see figure \ref{fig:NetChain} for a visual representation).
In this context, the dimension of the output of a given layer corresponds to the number of \textit{neurons} in that layer. Specifically, if there are $n^{(i)}$ neurons at layer $i$ in the network, the linear map from the $i^{\text{th}}$ to the $(i+1)^{\text{th}}$ layer is specified by the transformation
$
\bm{v}^{(i+1)}=M^{(i+1,i)}\bm{v}^{(i)}+\bm{b}^{(i+1)},
$
with {\it weight} matrix $M\in\mathbb{R}^{n^{(i+1)}\times n^{(i)}}$ and {\it bias} vector $\bm{b}\in\mathbb{R}^{n^{(i+1)}}$.

The two activation functions used in our network are an element-wise SoftSign function in the hidden layer $\text{SS}_i(\bm{v})\equiv v_i/(1+|v_i|)$,
where $v_i$ is the $i^{\text{th}}$ component of the Real vector $\bm{v}$, and the Softmax function 
$ \text{SM}_i(\bm{v})\equiv e^{-v_i}/\sum_{j}e^{-v_j}$,
in the output layer. In this way, the neural network can be summarised as the map
\begin{equation*}
\begin{split}
&f_{\text{net}}\text{: }\mathbb{R}^{n^{(1)}}\rightarrow \mathbb{R}^{n^{(3)}}; \quad \bm{v}^{(1)}\mapsto \bm{v}^{(3)},\\
    &f_{\text{net}}\left(\bm{v}^{(1)}\right)=\text{SM}\left[M^{(3,2)}\text{SS}\left(M^{(2,1)}\bm{v}^{(1)}+\bm{b}^{(2)}\right)+\bm{b}^{(3)}\right]\,.
\end{split}
\end{equation*}
The choice $M^{(3,2)}\in\mathbb{R}^{2\times n^{(2)}}$ ensures that the outputs are 2-dimensional vectors. The Softmax function, in turn, converts this into a vector $(v^{(3)}_1,v^{(3)}_2)^{\mathrm{T}}$ with $v^{(3)}_1+v^{(3)}_2=1$. Here, $v^{(3)}_1$ corresponds to the network confidence that the input is in an ETH phase, while $v^{(3)}_2$ gives the corresponding confidence level for the MBL phase.

In training the network on the nearest-neighbour model, we assume that systems with disorder sampled from $[-0.25,0.25]$ obey the ETH and map to $(1,0)$, while those with disorder sampled from $[-12,12]$ are in the MBL phase and map to $(0,1)$. These values are known to be well-inside the ETH (MBL) regions and we train our neural network at these disorder samplings, only. This scheme, arising from a training method that includes cross-validation, dropout regularisation, weight-decay, and lack-of-confidence penalties (see Supplementary Material), produces a neural network that can effectively classify entanglement spectra as being in the ETH or MBL phase, while systematically avoiding both over-fitting and classification agnosticism.\\

\begin{figure}[htb!]
     \centering
     \begin{subfigure}[b]{0.23\textwidth}
         \centering
         \includegraphics[width=\textwidth]{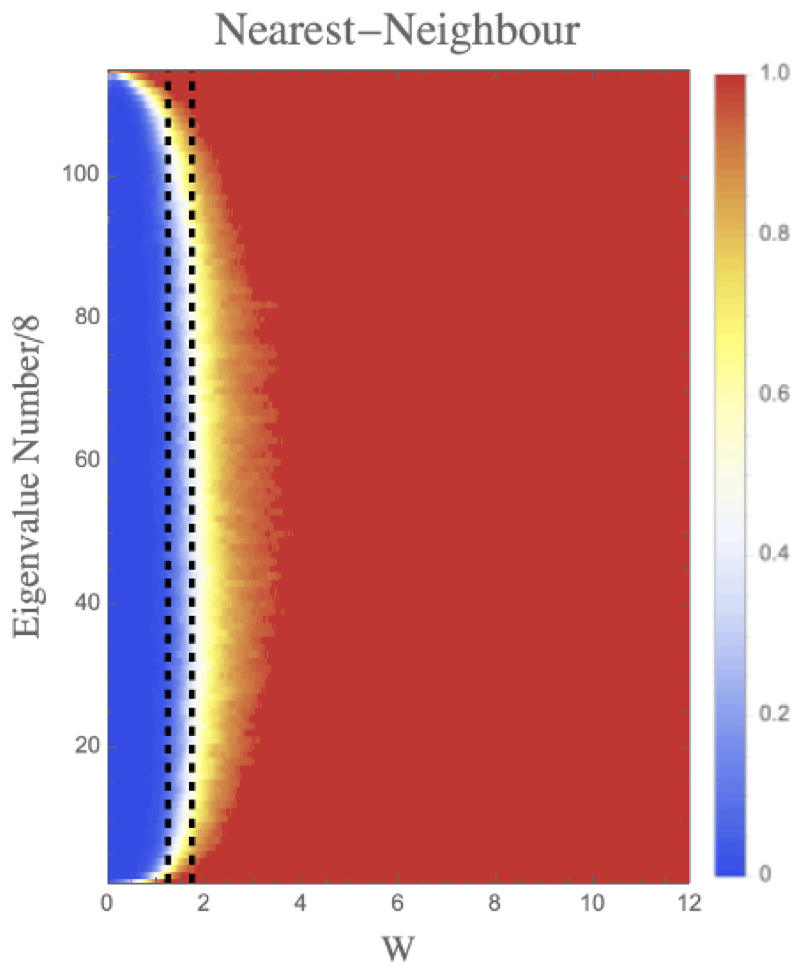}
         \caption{}
         \label{fig:NNPD}
     \end{subfigure}
     \hfill
     \begin{subfigure}[b]{0.23\textwidth}
         \centering
         \includegraphics[width=\textwidth]{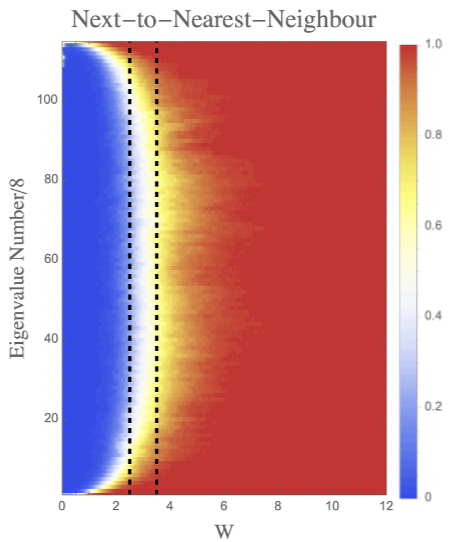}
         \caption{}
         \label{fig:NNNPD}
     \end{subfigure}
      \begin{subfigure}[b]{0.233\textwidth}
         \centering
         \includegraphics[width=\textwidth]{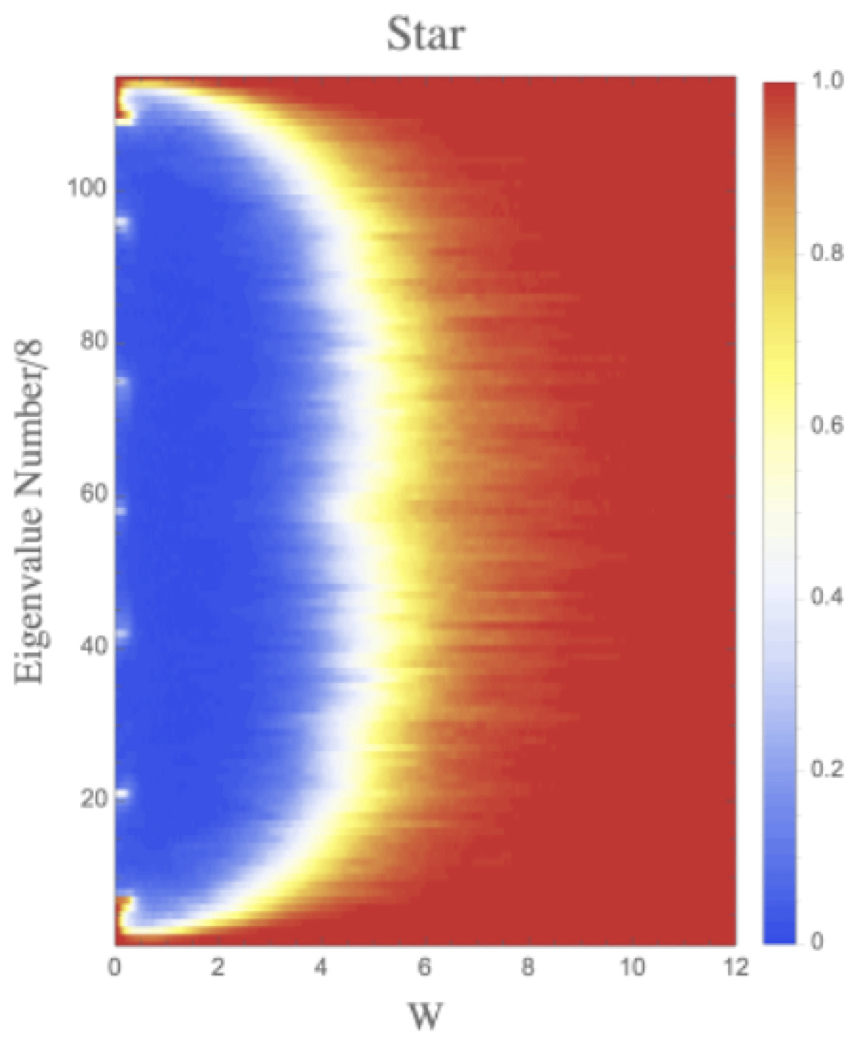}
         \caption{}
         \label{fig:STARPD}
     \end{subfigure}
     \hfill
      \begin{subfigure}[b]{0.233\textwidth}
         \centering
         \includegraphics[width=\textwidth]{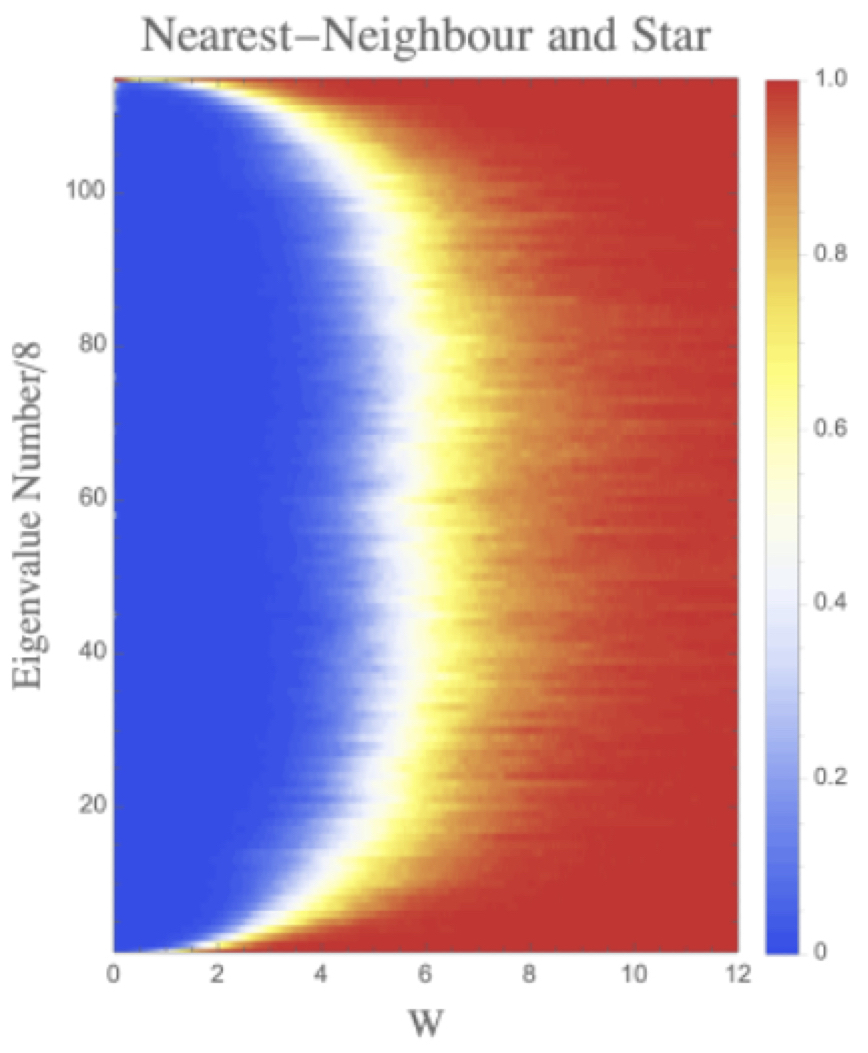}
         \caption{}
         \label{fig:NNSTARPD}
     \end{subfigure}
     \hfill
        \captionsetup{justification=raggedright}
        \caption{Phase diagrams for 48 disorder realisations of $N=12$ site systems with different graph topologies and disorder steps   $\Delta W=0.01$. All data is generated from the same 48 disorder realisations. In all diagrams, the data is grouped into bins of 8 consecutive energy eigenvalues and averaged inside each bin. We choose this binning over the $\epsilon$-binning as we do not claim to know anything about the distribution of the energy eigenvalues of all the topologies. The phase diagrams display the ETH/MBL transition for
        % and, for instance, the $\epsilon$-binning becomes a poorly defined procedure when there are no normalised-shifted energy eigenvalues inside a given $\epsilon$-bin (in other words, when a bin is empty). 
        (a) the nearest-neighbour system (b) the next-to-nearest-neighbour system, (c) the star graph, and (d) a superposition of the nearest-neighbour and star graphs. Dashed lines in (a) and (b) approximately correspond to regions exhibiting a sharp drop in the ratio of the system size, $N$, to the BKT correlation length  $\xi_{\text{BKT}}$ at $N/\xi_{\text{BKT}}\approx 0$ \cite{PhysRevE.102.062144}.}
        \label{fig:PDs}
\end{figure}

%\section{\label{sec:Results}Results}
\emph{\label{sec:Results}V. Results.---} To produce phase diagrams for a given system, we consider 48 unique disorder realisations at $W=1.0$ and scale the disorder at each site, for each realisation, by multiplying their values by the disorder value of interest. The entanglement spectra of each realisation and each eigenenergy is classified by the neural network at interval disorder values and these results are averaged either by grouped eigenenergy index, or by a  shifted-normalised eigenenergy binning. However, as we focussed on Hamiltonians with interaction graphs distinct from nearest-neighbour topology, we adopt the former since it  is robust against degenerate eigenenergies. The resulting phase diagrams are presented in figure \ref{fig:NNPD}. We note that the shape of the mobility edge in the nearest-neighbour case is in excellent agreement with existing literature for this case. This is particularly striking when considering the shifted-normalised eigenenergy binning \cite{PhysRevB.95.245134}.

We find that the ANN is also able to probe the phase boundary. To do so, we make the assumption that all data having MBL-classification confidence $p\geq0.9$ ($p\leq0.1$) are MBL (ETH) spectra, and are assigned a value of 0. On the other hand, spectra with $0.1<p<0.9$ are transition spectra, and are assigned a value of 1 \cite{PhysRevB.95.245134}. With these criteria and the same averaging over 48 realisations as was used in producing the phase diagram, we can effectively isolate the mobility edge between the ETH and MBL phases (see Supplementary Material).

We are now in a position to return to the question posed in the Introduction and turn our network, trained on the disordered nearest-neighbour chain, loose on more complex interaction graphs to see if is able to classify entanglement spectra in systems for which the neural network has had no training. As motivated in section {\bf III}, we consider three prototypical models, specified by their adjacency matrices, as representatives of broader classes of quantum networks. Applying precisely the same procedure as was done to produce figure \ref{fig:NNPD}, the network produces phase diagrams \ref{fig:NNNPD}-\ref{fig:NNSTARPD}.
At this point, there are several key observations to make:

%---The qualitative shape of the transition region associated with \ref{fig:NNPD} is in excellent agreement with known results [{\bf refs}]. This agreement becomes especially apparent when the bins used are based on the shifted-normalised eigenenergies (see, again, the Supplementary Material for more details).

---As the connectivity between sites is increased, we observe that the ETH region is 'pushed out' to larger disorder values. Heuristically, this is to be expected, as the coupling to the part of the system acting as the thermal bath is increased with increased connectivity. Along these lines, there are studies questioning and debating the existence of the MBL phase for long-range interactions and so our results sound very plausible \cite{PhysRevB.99.224203, PhysRevX.7.041021, PhysRevLett.113.243002}.
There are also direct comparisons, \cite{PhysRevE.102.062144}, of the Heisenberg and $J_{1}\!\!-\!\!J_{2}$ models whose breakdowns of the ETH appear to agree with the results found here (see dashed regions in figures \ref{fig:NNPD} and \ref{fig:NNNPD}), and these provide justification of the scaling of the ETH region found by the neural network in \cite{PhysRevB.95.245134}.

---Finally, in the case of the star graph (figure \ref{fig:STARPD}) and the superposition of nearest-neighbour and star networks (figure \ref{fig:NNSTARPD}) we see further expansion of the ETH region, as expected. 
Interestingly, the transition region is essentially the same in the two graphs, thus suggesting that connectivity is much more prominent than clustering in determining the MBL transition.
Additionally, in the system with star graph connectivity, eigenenergy degeneracy that we had not anticipated
%(but nonetheless that should exist due to the existence of an additional symmetry in the star system - external leg swapping at very-low disorder) 
was found at very-low disorder values and, as there is no `canonical' ordering of these degenerate eigenvalues, the entanglement spectrum is ill-defined. This manifests in the appearance of small `uncertainty bubbles' in the very low disorder region of figure \ref{fig:STARPD}. This degeneracy is removed by nearest-neighbour couplings as evidenced by the lack of these uncertainty bubbles in figure \ref{fig:NNSTARPD}.

%\section{\label{sec:Conclusions}Conclusions and Discussion}
\emph{\label{sec:Conclusions} VI. Conclusions and Discussion ---}
We have shown that a neural network, trained to recognize the ETH/MBL transition by means of the entanglement spectra in a prototype example, can efficiently detect the same kind of transition on new models \textit{without further retraining}.

This result is particularly relevant in promoting neural networks to the status of effective tools in investigations of MBL/ETH transitions. 
Indeed, confronted by a new model, one often does not know a disorder strength for which the model can be safely considered to be in the ETH (MBL) phase, hindering both training as well as subsequent results obtained by the neural network.

Together with the results of \cite{PhysRevB.95.245134}, showing that a \textit{simple} neural network detects the qualitative features of an MBL phase diagram \textit{of a model used to train the  network}; we have shown here that ANNs should be considered a powerful tool to perform \textit{preliminary investigations} of the phase diagram of a given quantum many-body system.
These investigations can (and should \cite{PhysRevB.100.224202}) be confirmed, and made more precise, at a later stage by more conventional (and computationally demanding) numerical tools \cite{PhysRevB.91.081103, PhysRevLett.125.156601}.

More speculatively, one can ask what is the ultimate reason for the excellent performances of neural networks on models different from the training model.
We believe that this is yet another manifestation of random matrix theory universality: in the ETH phase, typical eigenstates are supposed to be almost random while in the MBL phase the eigenstates should exhibit highly model-dependent features. 
Hence, we believe that the neural network learns how to distinguish universal from non-universal features and this should be the key to its performance while changing models.

Our work can be extended in several ways.
Allowing that it is an excellent tool for preliminary investigations, a natural question is whether it is possible to feed the neural network with simpler data than the entanglement spectrum, while retaining its efficiency in new models.
For example, \textit{how does the network perform working with the energy spectrum?} \cite{Kausar_2020}. 
At the physics level, our results show that the connectivity of the graph strongly affects the transition region. It would be important to make more robust and precise these findings, through more conventional techniques.
We plan to address all these points in the near future.

\emph{Acknowledgments.---}
DR acknowledges the support by the Institute for Basic Science in Korea (IBS-R024-Y2 and IBS-R024-D1). JM is supported in part by a Simons Associateship at the ICTP, Trieste. CB acknowledges support from the South African Research Chairs Initiative of the NRF. Computations were performed using facilities provided by the University of Cape Town’s ICTS High Performance Computing team. All graphics were produced by Wolfram Mathematica version 12.0.0.

\bibliography{main}% Produces the bibliography via BibTeX.

\clearpage
\setcounter{section}{0}
\setcounter{figure}{0}
\setcounter{equation}{0}

\appendix

\section{Supplementary Material for ``Neural Networks as Universal Probes of Many-Body Localization in Quantum Graphs"}

Appreciating that the machine learning tools that are at the core of this letter may not be entirely familiar to the bulk of its readership, and in the interest of self-containment, here we will provide an extended discussion of our neural network - its construction, architecture and training, as well as some of the terminology used - together with some of the preliminary results and checks that we undertook in order to increase our confidence in the new results reported in the letter. 

\subsection{\label{sec:ExtNN}I. Extended Neural Network Discussion}
\subsubsection{\label{sec:inpData}Input Data}
An $N$-site system of spin-1/2 particles has $2^N$ eigenstates. We will consider only those eigenstates with no net spin; {\it i.e.} $S_{\text{tot}}^z=0$. Consequently, we will enforce only even values for $N$ so that the number of energy eigenstates under consideration is $N \choose N/2$. When partitioning the system, we only ever split the system into equal halves ($N_A=N/2=N_B$), and trace out the $B$-subsystem. The constraint of zero net spin then enforces that $N$ be divisible by 4. Following \cite{PhysRevB.95.245134}, we also train the network on entanglement data. The entanglement (or modular) Hamiltonian \cite{Li:2008kda}, $H_e$, is defined through the reduced density matrix, $\rho_A$, obtained by tracing out the $B$-subsystem, as
\begin{equation}
    e^{-H_e}\equiv\rho_A=\text{Tr}_B\ket{\Psi}\bra{\Psi}.
\end{equation}
In this context, each energy eigenvalue is associated to an entanglement spectrum of size $N/2\choose N/4$. Additionally, we will use only the smallest half of these values, since larger values give vanishingly small contributions to the negative-exponential. Thus, as input for our network we always take an ${N/2\choose N/4}/2$ length vector that corresponds to the smallest half of the entanglement spectrum associated with a given eigenenergy (which, itself, will vary with the strength of the disorder). Hereafter we use the terms ``input vector" and ``entanglement spectrum" interchangeably.
The final limitation on the data used in training is that we consider only the central 80\% of eigenergies for training; we remove the lower and upper 10\% of eigenergies as they can have behaviour that strongly deviates from that of the central region, as is evidenced by figure \ref{fig:NNPD} in the main text, where we see that, for low disorder, the low energy states can exhibit MBL-like behavior, \cite{PhysRevB.95.245134}.

\subsubsection{\label{sec:NNTrainArch}Training Network Architecture}
\begin{figure*}[htb!]
    \centering
    \includegraphics[scale=0.6]{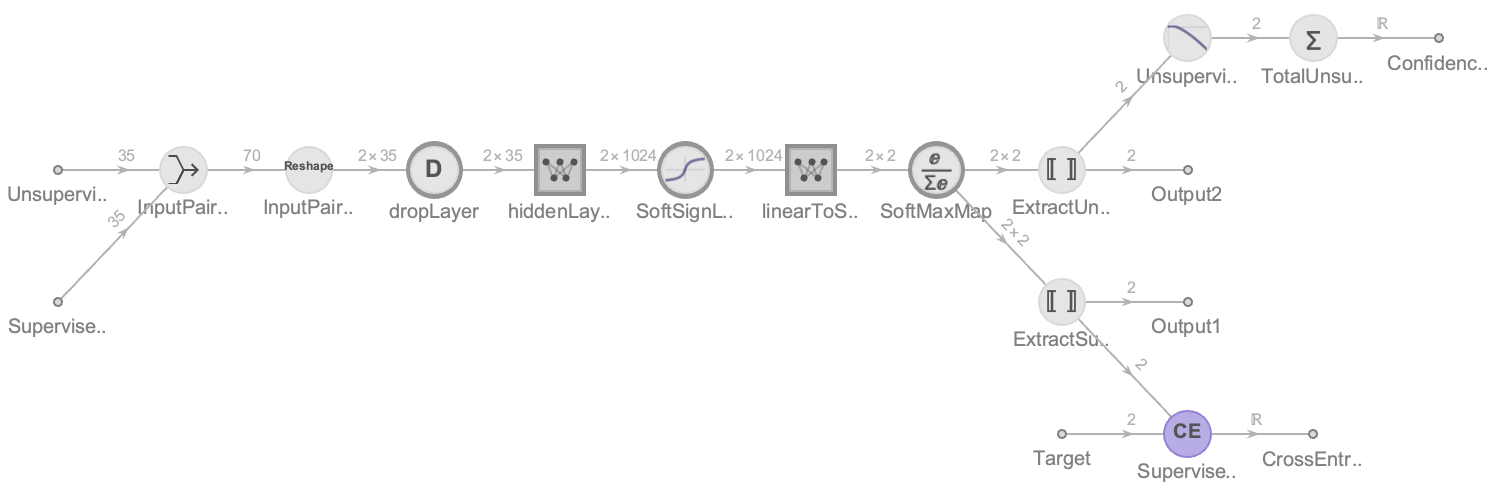}
    \captionsetup{justification=raggedright}
    \caption{The full network structure used for training for $N=16$ input sizes. Notice that there are two inputs (far left), two outputs (far right, middle), and two cost functions (far right, top and bottom). Of the larger nodes, nodes 4-7 constitute the part of the network we extract once training is complete.}
    \label{fig:NetGraph}
\end{figure*}

The training network architecture is a bit more complicated than that of the extracted network map described in the letter. To elaborate on this, we will make use of figure \ref{fig:NetGraph} which describes a training methodology built on four key attributes: a \textit{confidence penalty}, \textit{dropout regularization}, \textit{weight decay}, and \textit{cross validation}. \\

We will explicitly consider the neural network used to classify an $N=16$ site system. The only difference between this network and the $N=12$ site network is in the input sizes - the same number of neurons are used at the hidden layer and thereafter. 

\textbf{Confidence Optimization}: For a $N=16$ system, the lower half of the entanglement spectrum (the network input) has size ${8\choose 4}/2=35$, whereas the training network in \ref{fig:NetGraph} has two of these as inputs. This is because, at every training step, two input vectors are given - the first being taken from the disorder region of assumed phase (either $0.25\rightarrow\text{ ETH}$ or $12.0\rightarrow \text{ MBL}$) which is accompanied by its target vector (being $(1,0)$ and $(0,1)$, respectively) which is called \textit{supervised} input, and the second being a vector from a region of intermediate disorder $W\in(0.25,12.0)$ where no target vector is specified which is called \textit{unsupervised input}. The first two larger nodes in \ref{fig:NetGraph} simply reshape this data into a form the network can process. At the remaining large nodes are identical copies of the network through which both inputs are modified. For the supervised input we use the conventional cross-entropy cost function \cite{Mehta:2018dln},
\begin{equation}
    \text{Sup. Cost}(\hat{f}, f_\text{net}) = -\sum_{\bm{v}\in\text{TD}}\sum_{i=1}^2 f_{\text{net},i}(\bm{v}_i)\ln \hat{f}(\bm{v}_i),
    \label{eq:sup}
\end{equation}
where $f_\text{net}$ is the network map, $\hat{f}$ is the target map and the first sum is over the input vectors in the $W=0.25$ or $W=12.0$ regions (`Targeted Data'). The unsupervised input, not having a target vector to take the cross-entropy with is used to enforce the confidence penalty by noticing that the function $C(x)=x\log x$, with domain $x\in[0,1]$ has global minima at values of 0 and 1. Thus we use the Shannon entropy as a cost function for the unsupervised data \cite{6773024, 6773067}
\begin{equation}
    \text{Unsup. Cost}(f_\text{net}) = -\sum_{\bm{v}\in\text{ID}}\sum_{i=1}^2 f_{\text{net},i}(\bm{v}_i)\ln f_{\text{net},i}(\bm{v}_i),
    \label{eq:unsup}
\end{equation}
where the first sum is over the `Intermediate disorder Data'. The inclusion of this cost function penalizes the neural network for finding values farther from 0 and 1, pushing the network towards being as confident as possible in the intermediate disorder region.  One could choose to moderate the influence of this penalty by multiplying through \eqref{eq:unsup} by a hyperparameter, but investigation finds that this value is optimal for values near 1, and so we do not explicityly investigate the effects of this hyperparameter here \cite{PhysRevB.95.245134}. Considering a cost function that combines \eqref{eq:sup} and \eqref{eq:unsup}, such a cost function would push the netword towards classification of (assumed) definite-phase states and unknown-phase states with minimal agnosticism. 

\textbf{Dropout Regularization}: To avoid overweighting particular neurons, we make use of dropout regularization; at every training step some of the neurons are not trained at all. We do not enforce that a specific number of neurons not be evaluated at each step, rather we enforce that every neuron has a probability of 0.5 of being dropped at every training step.

\textbf{Weight Decay}: At any point during training the weights within a linear map may be sporadically transformed into nonzero values to produce anomalous overfittings. To avoid this, we implement weight decay of the transformation matrix in the hidden layer. To be precise, we make use of a stochastic gradient descent training algorithm that updates the weights and biases of the linear layers at each training step by randomly selecting a subset of the training data and minimizing the error on \textit{that} subset of the training data, that is we minimize the error (cost) on
\begin{equation}
    A_{i+1} = A_i-\lambda \frac{\partial f_{\text{cost}}}{\partial A_i},
    \label{eq:SGD}
\end{equation}
where $A_i$ are the weights and biases at training step $i$, $\lambda$ is the learning rate hyperparameter, and $f_\text{cost}$ is the cost function, for a subset of the total training data. The effect of including weight decay is to add a decay term $-\mu A_n$ to \eqref{eq:SGD}, where $\mu$ is the weight decay hyperparameter. The inclusion of this term in \eqref{eq:SGD} sees a corresponding change to the cost function, where we must also now include a $+\mu|A|^2$ (for the $l_2$-norm $|\cdot|$). We choose to apply this weight-decay procedure only to the hidden layer weight matrix. We used values of $\lambda=10^{-4}$ and $\mu = 1$ for training, as is done in \cite{PhysRevB.95.245134}.

\textbf{Cross-Validation}: Finally, before training starts all the training data is randomly partitioned into two sets if equal cardinality. One set is used for training, while the other (validation) set is used to ensure overfitting is minimised by comparing how the updated network affects the error measured on a validation set - a general neural network should be attempt to minimize the error on both the training data and the validation data.

Including all considerations thus far, the final cost function is
\begin{equation}
\begin{split}
    \text{Cost}&(\hat{f},f_{\text{net}})=-\sum_{\bm{v}\in\text{TD}}\sum_{i=1}^2 f_{\text{net},i}(\bm{v}_i)\ln \hat{f}(\bm{v}_i)\\
    &-\sum_{\bm{v}\in\text{ID}}\sum_{i=1}^2 f_{\text{net},i}(\bm{v}_i)\ln f_{\text{net},i}(\bm{v}_i)
    +\mu |M|^2,
\end{split}
\end{equation}
where $M$ is the weight matrix of the hidden layer, and this is the cost function used while training.

\begin{figure}[hbt!]
     \centering
     \begin{subfigure}[b]{0.46\textwidth}
         \centering
         \includegraphics[width=\textwidth]{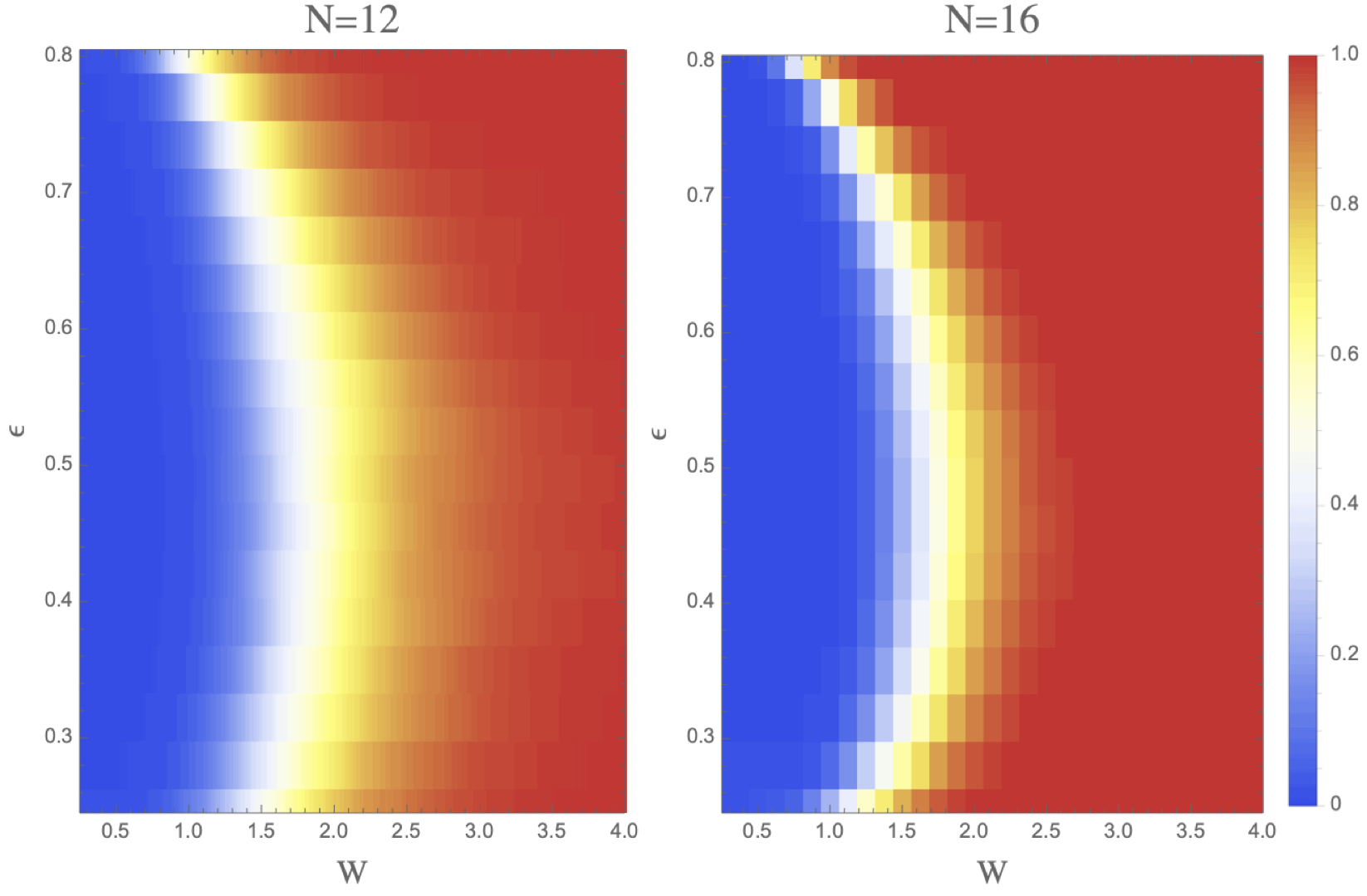}
         \caption{}
         \label{fig:Supp12vs16NNPD}
     \end{subfigure}
     \hfill
     \begin{subfigure}[b]{0.46\textwidth}
         \centering
         \includegraphics[width=\textwidth]{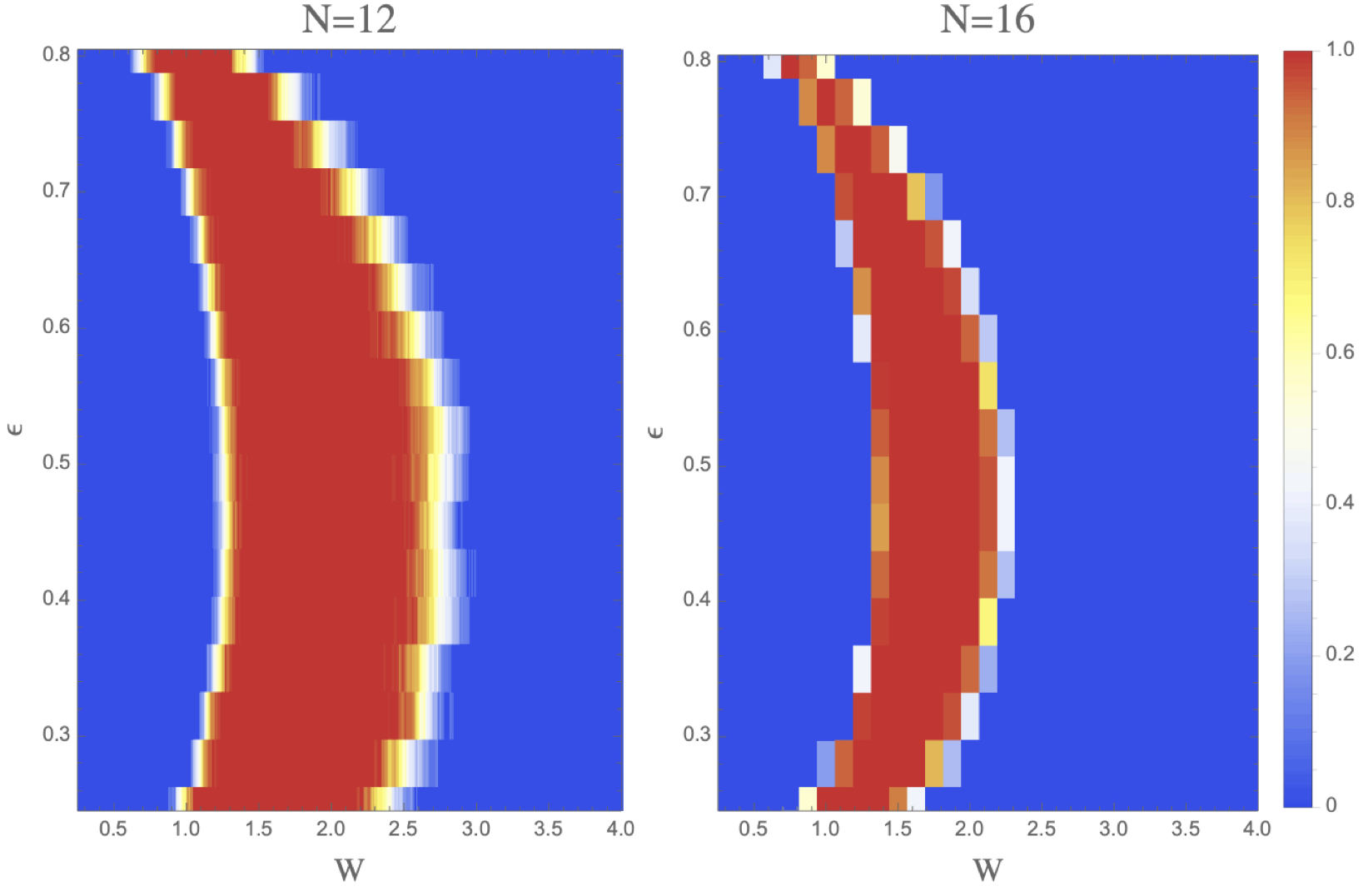}
         \caption{}
         \label{fig:Supp12vs16NNTP}
     \end{subfigure}
        \captionsetup{justification=raggedright}
        \caption{Phase diagrams and transition region probing (using the method described in the supplementary text) for different system sizes averaged over 48 disorder realisations for nearest-neighbour systems. The $N=12$ site system has values computed in disorder steps of $\Delta W=0.01$; the $N=16$ site system has $\Delta W=0.125$. All figures run from $W=0.25$ to $W=4.0$. (a) The phase diagrams for an $N=12$ site system (left) and an $N=16$ site system (right). (b) The associated transition region probes associated with the phase diagrams in (a).}
        \label{fig:NNPhaseDiagrams}
\end{figure}

The training method (stochastic gradient descent) and associated additional techniques (cross-validation etc,) described above are in no way unique to this work and there exist numerous resources. We point the reader to \cite{Mehta:2018dln} and references therein. Furthermore, the reader may note a strong similarity with this work and the work done in \cite{PhysRevB.95.245134}; this is intentional. We, however, make the use of the SoftSign activation function in the hidden layer instead of the Rectified Linear Unit activation function, as empirical investigation showed that the SoftSign function produced the best classification confidence and a minimal transition region (thinnest red region in, for instance, figure \ref{fig:Supp12vs16NNTP}).

\subsubsection{\label{sec:NNTrainResults}Training Results}
The results of the trained network are presented in table \ref{tab:NNTrainRes}. These results characterize the ability of the networks to recognize ETH/MBL type systems, with greater accuracy achieved when greater system size is used.\\

\begin{table}[h]
\captionsetup{justification=raggedright}
\caption{\label{tab:NNTrainRes}%
The trained network's classification results. System size refers to the number of sites the system has (i.e.the value of $N$). The data type specifies for which sort of training data we are calculating statistics; The mean is the average confidence of the network in the data being of the type specified by data type, and the standard deviation is the standard deviation of the mean. The data set sizes before partitioning for cross-validation were $37050$ for $N=12$ and $514850$ for $N=16$ (excluding intermediate disorder data).
}
\begin{ruledtabular}
\begin{tabular}{lcdr}
\textrm{System Size}&
\textrm{Data Type}&
\multicolumn{1}{c}{\textrm{Mean (\%)}}&
\textrm{Std Dev. (\%)}\\
\colrule
12 & ETH & 99.35 & $1.3\times10^{-1}$\\
12 & MBL & 99.99999 & $3.0\times10^{-5}$\\
16 & ETH & 99.933 & $1.4\times10^{-2}$\\
16 & MBL & 100.00 & $<10^{-10}$\\
\end{tabular}
\end{ruledtabular}
\end{table}
In addition to the results specified in table \ref{tab:NNTrainRes}, we also mention the \textit{least} confident classification value for $N=12$: 96.9428\% in the ETH regime and 99.9996\% in the MBL regime. Similarly, these \textit{worst} confidence values for $N=16$ were 99.78\% and 100.00\%, respectively. Assuming all states with classification confidence $90\%<$ are \textit{in} that state, the network classifies 100\% of the training and validation data successfully post-training. These results also show that the networks are particularly good at classifying MBL phases.

% One can attempt to probe what the ANN has learned by asking it what `perfectly' ETH or MBL input data should look like. To do this, one takes an already trained network and holds it fixed as an unchanging map, and prepends to it an additional layer before it reaches the map. One can then use stochastic gradient descent and enforce two limitations; \textit{nonnegativity} (that a physically realisable entanglement spectrum must be nonnegative) and \textit{ordering} (that an entanglement spectrum must be ordered by numeric value from smallest to largest). Taking data that was poorly classified (taking on values of $0.4<v_i<0.6$) as our training set and enforcing the target output be either (1,0) always or (0,1) always, one may probe what the network `knows' \cite{PhysRevB.95.245134}. The results from such an analysis are presented in figure ... ***we can add the dreaming plots for the main paper - the Mathematica version on the HPC is old and doesn't support the functionality I used when implementing on my laptop. So, hopefully, we should be able to show the dreaming results in a meaningful way later (as well as N=16 data)***

\subsection{\label{sec:ExtResults}II. Extended Results}
In the paper, we make use of the phase diagrams found by averaging over 48 unique realisations of an $N=12$ site system, and further grouping and averaging over 8 consecutive eigenvalues (see figure \ref{fig:PDs}). While this methodology is essential for the probing of new topologies whose eigenenergy structure and spacing is not assumed to be known, an alternative methodology that may be used is grouping by binned and shifted-normalised eigenergies,
\begin{equation}
    \epsilon=(E-E_{\text{min}})/(E_{\text{max}}-E_{\text{min}}),
    \label{eq:enshift}
\end{equation} 
where $E$ is a given eigenenergy, and $E_{\text{min(max)}}$ is the minimum(maximum) eigenenergy for a given disorder realisation at a given disorder value. The result of using this grouping methodology is presented in figure \ref{fig:Supp12vs16NNPD}.

As we are interested in the phase transition from ETH to MBL, it is instructive to highlight the region in which the network struggles to confidently classify entanglement spectra. To do this, we produce and classify 48 unique disorder realisations over the range $W\in[0.25,12]$, as before. However, once the network has classified the data, we assume all output 2-vectors having first entry $v_1<0.1$ are in the MBL phase and are set to 0, all $v_1>0.9$ are in the ETH phase and are \textit{also} set to 0, and the remaining data ($0.1\leq v_1 \leq 0.9$) are in the transition region and are set to 1. In this way we produce a heat map of the transition region. The process of averaging over realisations eigenergies (either by eigenenergy-grouping or $\epsilon$-binning) gives rise to yellow-tinted regions, as shown in figure \ref{fig:Supp12vs16NNTP}.

In particular, the shape of the $N=16$ phase diagram should be compared to \cite{PhysRevB.95.245134} and \cite{ALET2018498} after a purely superficial change in the aspect ratio of the image; where there is notable similarity in the both cases up to a re-scaling of the critical disorder values.

\begin{figure}[h!]
     \centering
     \begin{subfigure}[b]{0.23\textwidth}
         \centering
         \includegraphics[width=\textwidth]{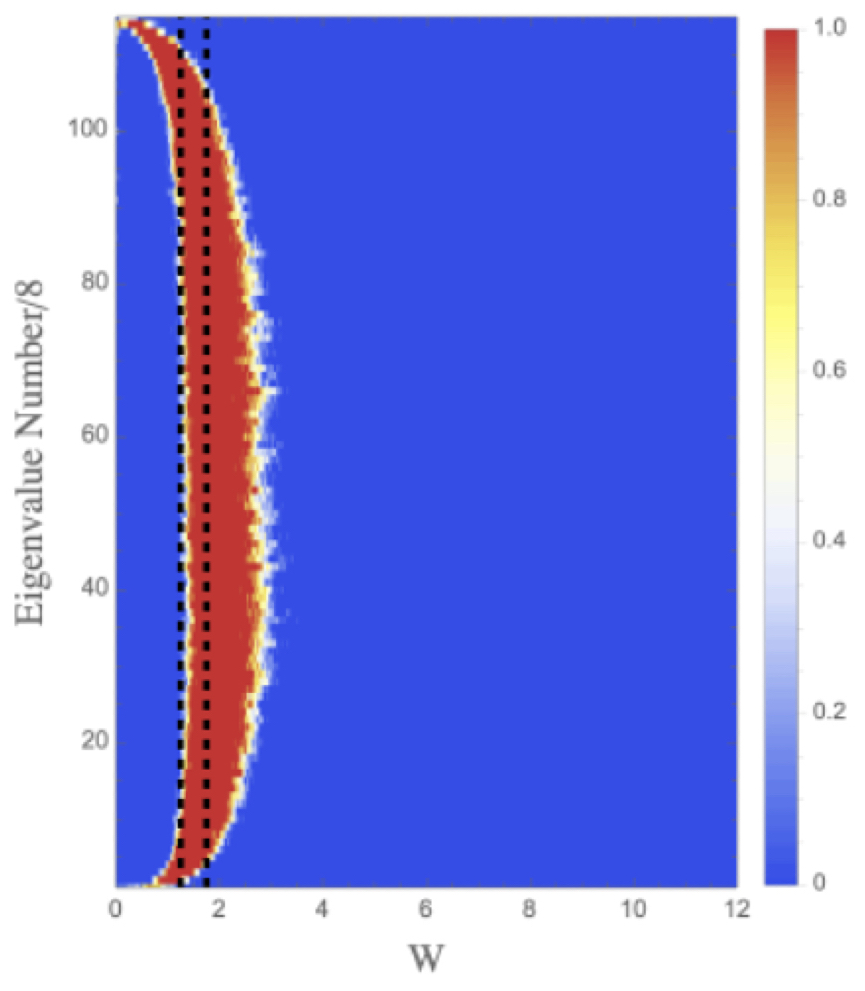}
         \caption{}
         \label{fig:SuppNNPD}
     \end{subfigure}
     \hfill
     \begin{subfigure}[b]{0.23\textwidth}
         \centering
         \includegraphics[width=\textwidth]{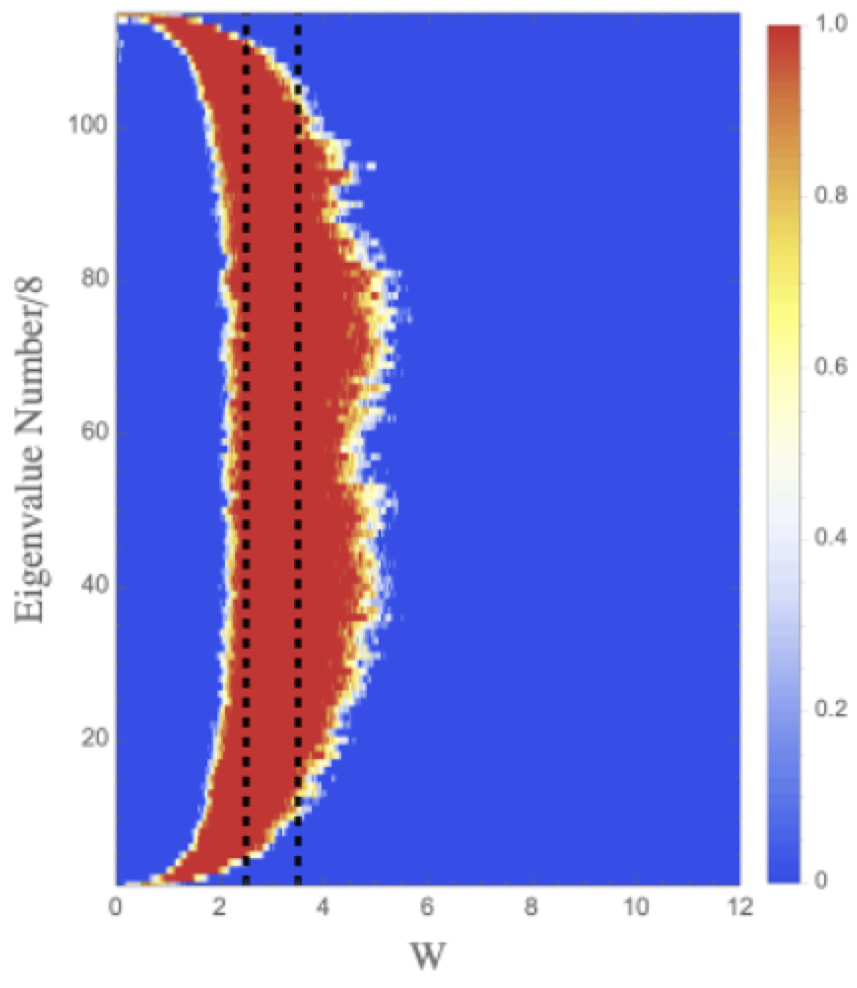}
         \caption{}
         \label{fig:SuppNNNPD}
     \end{subfigure}
      \begin{subfigure}[b]{0.23\textwidth}
         \centering
         \includegraphics[width=\textwidth]{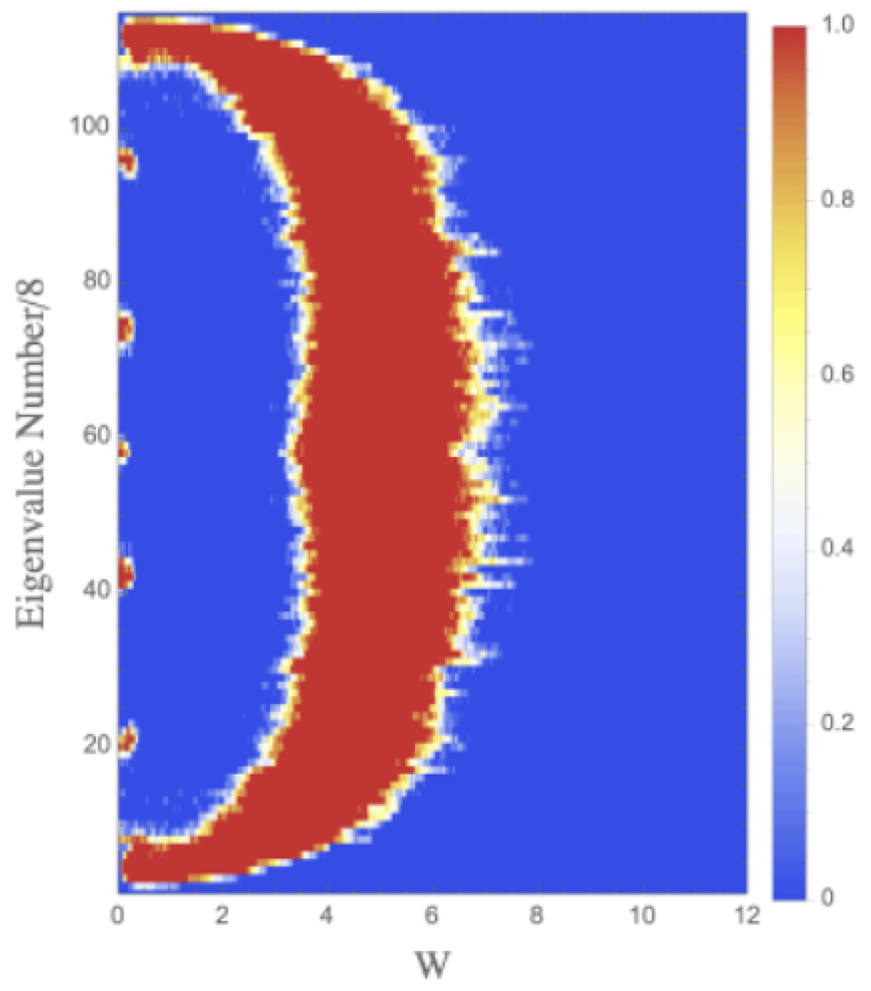}
         \caption{}
         \label{fig:SuppSTARPD}
     \end{subfigure}
     \hfill
      \begin{subfigure}[b]{0.23\textwidth}
         \centering
         \includegraphics[width=\textwidth]{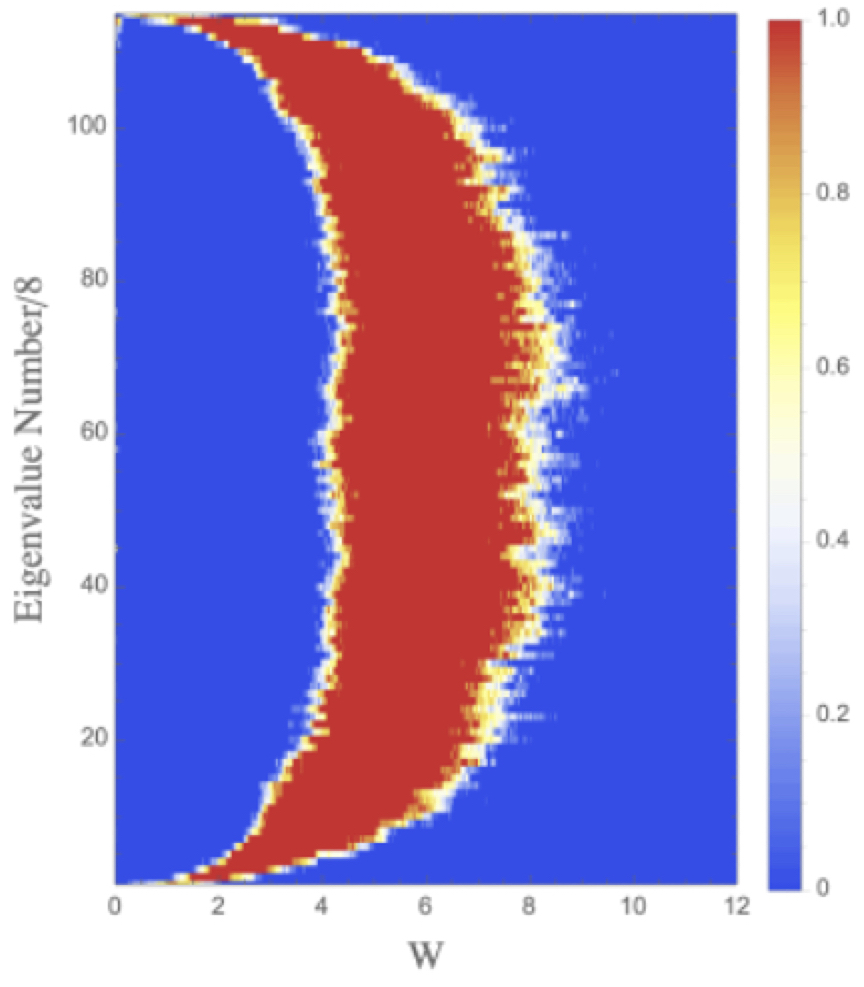}
         \caption{}
         \label{fig:SuppNNSTARPD}
     \end{subfigure}
     \hfill
        \captionsetup{justification=raggedright}
        \caption{The associated transition regions to the phase diagrams in figure \ref{fig:PDs} in the paper for 48 realisations of $N=12$ site systems having different topologies. In all diagrams, we group data into bins of 8 consecutive energy eigenvalues and average the data inside each bin.(a) Results for the nearest-neighbour system. (b) Results for the next-to-nearest-neighbour system. (c) Results for the star system. (d) Results for the bicycle-wheel system (superposition of nearest-neighbour and star systems). The dashed lines in (a) and (b) approximately correspond to regions in which the there is a sharp drop in the ratio of the system size, $N$, to the Berezinskii-Kosterlitz-Thouless correlation length  $\xi_{\text{BKT}}$ at $N/\xi_{\text{BKT}}\approx 0$ \cite{PhysRevE.102.062144}.}
        \label{fig:SuppPDMEL}
\end{figure}

We can probe the transition region in all topologies using either averaging methodology. The transition region probe for the phase diagrams in the paper (see figure \ref{fig:PDs} in the paper) is presented in figure \ref{fig:SuppPDMEL}. The small `bubbles' of high uncertainty at low disorder in figure \ref{fig:SuppSTARPD} may be of interest to the reader. These bubbles of uncertainty occur as a result of an energy degeneracy in the star system. As there is not a canonical way of ordering these degenerate eigenvalues, the notion of an entanglement spectrum becomes ill-defined and hence it does not make sense to try classify this region. Nonetheless, we include it so that the various topologies may be compared directly.

\end{document}